\documentclass[12pt,preprint]{aastex}
\usepackage{graphicx}
\usepackage{natbib}

\begin{document}

\renewcommand{\topfraction}{1}
\renewcommand{\bottomfraction}{1}
\renewcommand{\textfraction}{0}
\renewcommand{\floatpagefraction}{1.1}

\textheight 9.0in

\title{Trigonometric Parallaxes and Proper Motions of 134 Southern Late 
M, L, and T Dwarfs from the Carnegie Astrometric Planet Search Program}

\author{Alycia J.~Weinberger, Alan P.~Boss, Sandra A.~Keiser}
\affil{Department of Terrestrial Magnetism, Carnegie Institution for
Science, 5241 Broad Branch Road, NW, Washington, DC 20015-1305}

\author{Guillem Anglada-Escud\'e}
\affil{School of Physics and Astronomy, Queen Mary University of London, 327
  Mile End Road, London, E1 4NS}

\author{Ian B.~Thompson}
\affil{Carnegie Observatories, Carnegie Institution for Science, 813 
Santa Barbara Street, Pasadena, CA  91101-1292}

\author{Gregory Burley}
\affil{National Research Council of
  Canada, 5071 West Saanich Road, Victoria, BC V9E 2E7, Canada}

\begin{abstract}

We report trigonometric parallaxes for 134 low mass stars and brown dwarfs, of
which 38 have no previously published measurement and 79 more have improved
uncertainties. Our survey targeted nearby targets, so 119 are closer than 30
pc. Of the 38 stars with new parallaxes, 14 are within 20 pc and seven are
likely brown dwarfs (spectral types later than L0). These parallaxes are
useful for studies of kinematics, multiplicity, and spectrophotometric
calibration. Two objects with new parallaxes are confirmed as young stars with
membership in nearby young moving groups: LP 870-65 in AB Doradus
and G 161-71 in Argus. We also report the first parallax for the
planet-hosting star GJ 3470; this allows us to refine the density of its
Neptune-mass planet.  One T-dwarf, 2MASS J12590470-4336243, previously thought
to lie within 4 pc, is found to be at 7.8 pc, and the M-type star 2MASS
J01392170-3936088 joins the ranks of nearby stars as it is found to be within
10 pc. Five stars that are over-luminous and/or too red for their spectral
types are identified and deserve further study as possible young stars.
\end{abstract}

\keywords{astrometry -- stars: low-mass, brown dwarfs}

\section{Introduction}

Determination of the physical properties of low mass stars and brown dwarfs,
most importantly luminosities, depends upon having accurate
distances. However, these late-type objects were generally too faint for
inclusion in the all-sky Hipparcos survey.  Through the efforts of several
ground-based astrometric surveys, there are now hundreds of low-mass stars
with parallaxes
\citep[e.g.][]{Dahn:2002,Jao:2005,Costa:2006,Andrei:2011,Dupuy:2012,Faherty:2012,Dieterich:2014,Sahlmann:2014,ZapateroOsorio:2014}.
  Nevertheless, there are still many objects within 30 pc without
  well-measured distances.  These nearby, bright objects would be the best
  templates for studies of radii, atmospheric composition, metalicity, and
  other spectroscopic properties. In addition, low mass stars with excellent
  distances provide the templates for spectrophotometric distances to more
  distant stars.

In 2007, we began a long-term astrometric search for gas giant planets and
brown dwarfs orbiting nearby low mass dwarf stars \citep{Boss:2009}.  The
search employs a specialized astrometric camera, the Carnegie Astrometric
Planet Search Camera (CAPSCam), with a design optimized for high accuracy
astrometry of M dwarf stars.  Here we report our trigonometric parallaxes for
134 low mass stars. Of these, 38 have no previously reported measured parallax.

\section{Observations}

CAPSCam operates on the 2.5-m du Pont telescope at the Las Campanas
Observatory in Chile and is described in detail by \citet{Boss:2009}; its main
features for astrometry of low mass stars are briefly described here. CAPSCam
has no internal moving parts and employs an astrometric quality filter as its window
that is approximately z-band (865 nm with a bandpass of 100 nm). The field of
view is 6.7 arcmin on a side, with 2048$\times$2048 pixels each subtending
0$\farcs$196. A subarray, also known as the ``guide window,'' is arbitrarily
sizable and locatable and may be read out independently from the rest of the
field.  A bright target star is placed in the guide window, which is then read
out fast enough so the star does not saturate while the rest of the pixels
integrate on the reference stars; a mechanical shutter in front of the
entrance window ensures that the exposure time on the bright star remains, as
much as possible, commensurate with that on the full field. Thus, the camera
can achieve high dynamic range without excessive overhead. 

Target selection concentrated on southern M, L and T dwarfs closer than 20 pc
as known from either parallaxes or spectrophotometric distances. At the time
of the initial target selection ten years ago, distances and spectral
sub-types for many late type stars were lacking, so high proper motion stars
were included as well. The earliest spectral type included was M3, and the
majority of targets are spectral type M5.5 and later. In 2011, the target list
was updated to include all objects with spectral type later than M4, closer
than 12 pc, and south of declination $+$16$^\circ$. Stars must have I
magnitudes greater than $\sim$9 so as not to saturate the detector in the
minimum guide window exposure time of 0.2 s. The faintest objects we target
have I$\sim$18 so as to provide S/N$\sim$500 in a 120~s integration.

Our typical observing strategy is to place target stars brighter than
I$\sim$15 in the guide window.  Full field integration times are chosen to get
at least 6, and typically more like 25, well-exposed reference stars; the
number of reference stars for each field is given in Table \ref{tab:obs}. The
typical astrometric reference star for our fields has I$\sim$17 and can be as
faint as I$\sim$22.  The usual integration times are also given in Table
\ref{tab:obs}, although in some epochs, they were adjusted for seeing and
clouds.  At each epoch, we typically observe for an hour and thus obtain 20-40
images of the full field.  Targets are almost always observed within an hour
of transit, and given the long wavelength filter of the camera there is little
differential atmospheric refraction as a function of stellar spectral type.

The data for our parallaxes were collected from 2007-2014. The number of
epochs per source varies from 4, the minimum to obtain a parallax with
uncertainty estimates, to more than 20 for a few well-studied targets. The
number of epochs, and the start and end dates for the data, and time baseline
of the observations included in the parallaxes are given in
Table \ref{tab:obs}. We typically observe each star at least twice per
calendar year. The stars range in spectral type from M3 through T7, with the
bulk of the targets being late M-type.

\section{Data Reduction}

Details of CAPSCam astrometric data reduction may be found in
\citet{Boss:2009} and \citet{Anglada:2012} and are briefly summarized here following the
description in \citet{Weinberger:2013}.  For each epoch, the x and y pixel
positions of the brightest $\sim$100 stars (more in crowded fields) in the
field are found with a centroiding algorithm. Data from all epochs are
combined in an astrometric solution to derive the positions, proper motions,
and parallaxes of all the cross-matched stars in each target field.  The
astrometric solution is an iterative process. An initial catalog of positions
starts with the centroids from a chosen epoch transformed to sky coordinates
based on the coordinates of the target star and the known pixel scale. Next, a
transformation is applied to every other epoch's catalog to match the initial
catalog, and the apparent trajectory of each star is fit to a basic
astrometric model. The parallaxes for all objects are initialized to zero. The
initial catalog is updated with new positions, proper motions, and parallaxes,
and a subset of well-behaved stars is selected to be used as the reference
frame. The reference stars must be successfully extracted in every epoch and a
subset of at least 15, and more typically 30, is chosen that shows the
smallest epoch-to-epoch variation in their solutions. This process is then
iterated a small number of times.

In each iteration the individual parallax and proper motions of every star are
adjusted, so the mean parallax should stay at approximately zero. However, the
subset of reference stars do not necessarily have mean parallax of zero. At
any epoch, the position of a star has centroiding uncertainties, and for
distant stars, proper motion will take out all apparent motion of the star,
leaving positional residuals that are both positive and negative. Therefore,
although the true parallax to every star must be positive, we allow the fit
parallaxes to take on positive and negative values.

To assess the uncertainties on the measured parallax, we perform a Monte Carlo
where we fit the starting position, parallax, and proper motion in each
trial. Each trial draws random positions for each epoch based on the nominal
position determined from the iterative solution and its positional
uncertainty. If the $\chi^2$ of the parallax fit is $>$1, we add to every epoch's
uncertainties and re-fit until $\chi^2$ equals one. This additional uncertainty, or
positional jitter, may arise from any sources of systematic uncertainty. The
final parallax uncertainty is the standard deviation in the parallaxes of each
trial.

The final astrometric solution gives the motion of all the stars in the
field. However, these stars have parallactic motions that are all in the same
direction, since they are generated by Earth's motion. This introduces a small
bias, also known as a zero-point parallax offset, that must be removed to find
the absolute parallax.  

To find the zero point for each field, we estimate a photometric distance to
the brightest reference stars by fitting a Kurucz stellar model to cataloged
USNO-B1 magnitudes at B2, R2, and I \citep{Monet:2003} and 2MASS magnitudes
at J, H, and Ks \citep{Skrutskie:2006} and assuming each star is a
dwarf. Dwarf stars with fit Teff $<$ 4000 K are excluded. We average the difference
between our astrometrically determined (even if they are not statistically
significant) and photometric parallaxes to find the average bias and its
uncertainty and subtract it from our relative parallaxes and propagate the
uncertainty.  We cannot make a comparable zero-point proper motion correction
because so few stars as faint as our reference stars have measured absolute
proper motions.   For 18 of our fields, we were unable to compute a zero point
correction due to a combination of reference stars that were too cool and/or
faint to be fit well. However, inspection of Table \ref{tab:results} shows that
our typical zero point correction is small ($<$ 1 mas) and that the average
correction across the stars for which they were computed is -0.09 $\pm$ 0.43
mas. Therefore, for these 18 objects, we assumed no zero point correction and an
additional uncertainty of 0.4 mas.

\section{Results}
\begin{figure}[ht]
\centering\includegraphics[width=4.5in]{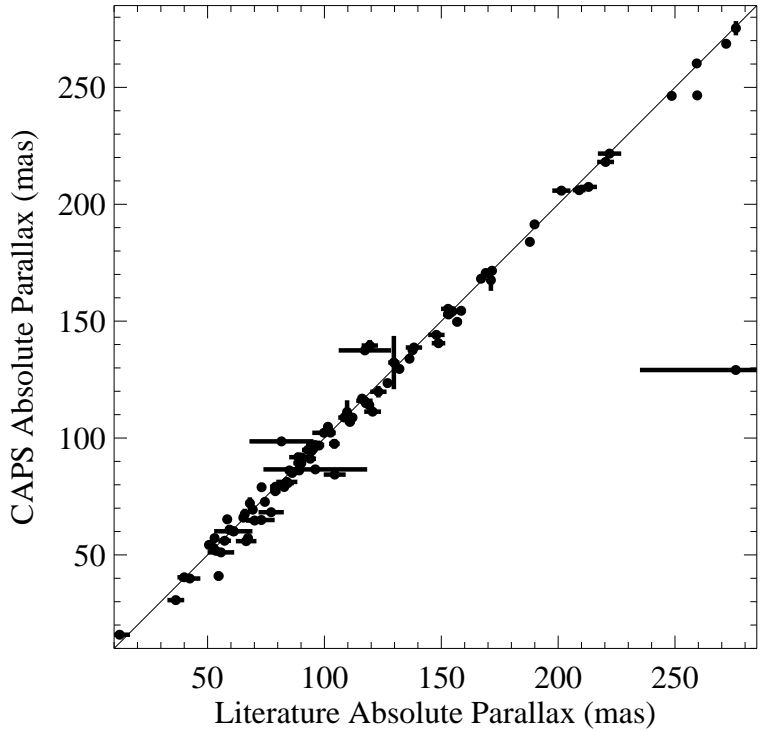}
\caption{\small A comparison of parallaxes for 95 of the CAPS targets for which literature values
  exist. The diagonal line is drawn as a guide and is not a fit. The obviously discrepant
  point is 2MASS J1259-4336, and is discussed in Section \ref{s:J1259}. Not shown is GJ
  406, the closest star in our sample, whose parallax is 413 mas.\label{fig:compare}}
\end{figure}

Table \ref{tab:results} lists the relative parallaxes, relative proper motions,
zero-point parallax corrections, and final absolute parallaxes for all our
targets as well as previously published trigonometric parallax values from the
literature. Figure \ref{fig:compare} compares our absolute parallaxes with
published parallaxes from other work.

For 79 of the 96 stars with previously published parallaxes, our measurements have
lower uncertainty. In general there is very good agreement between ours and previous
measurements; only 12 of the 96 disagree by more than 3 $\sigma$ (of the less
accurate measurement), and for 8 of these 12, the difference in parallax is $<$5\%.
The remaining four discrepant sources are explained in more detail, below.

A formal least-squares fit to the published trigonometric parallaxes versus ours gives a
slope of 0.988 $\pm$ 0.003, i.e., the CAPSCam parallaxes are slightly low compared to
published values -- an average of 2.9 mas low. However, the $\chi^2$ of this fit is poor,
which suggests that either the literature uncertainties, our uncertainties, or both, are
underestimated. Note also that this comparison includes the poor matches addressed below.

\begin{figure}[htb]
\centering\includegraphics[angle=90,width=5.95in]{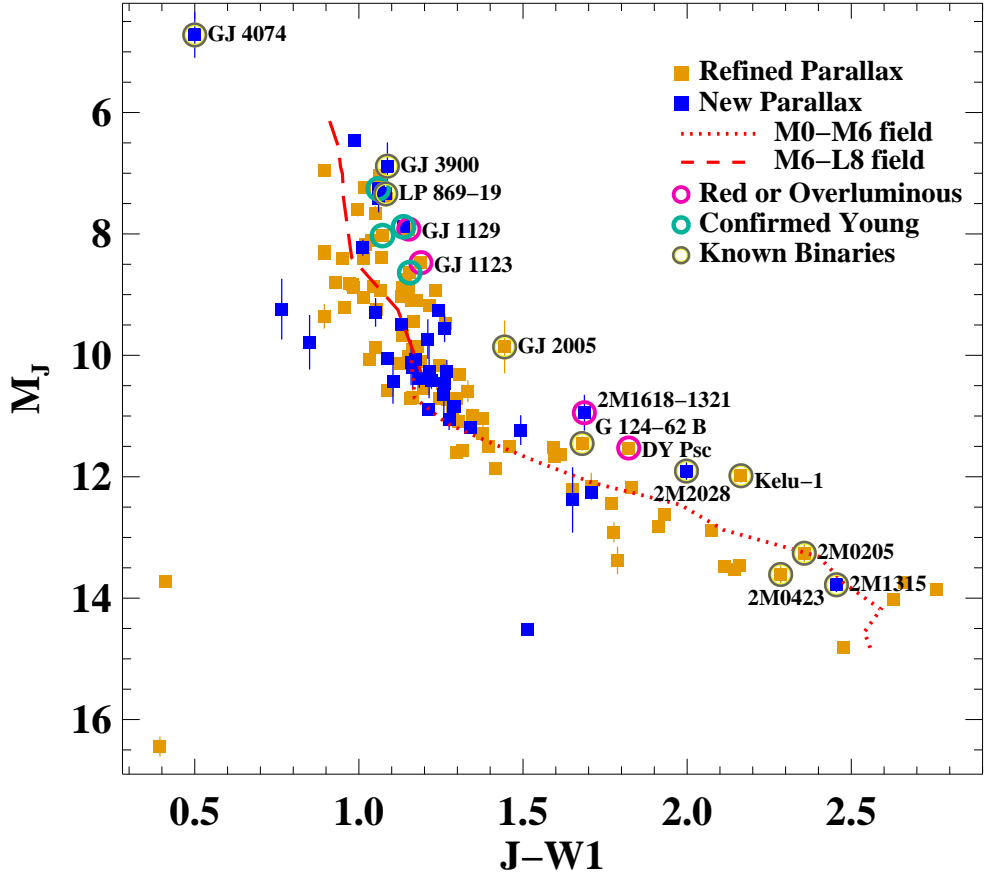}
\caption{\small The J-W1 (WISE Band 1) versus $M_J$ color-magnitude diagram including
  all the stars for which we obtained parallaxes. Objects without previously
  published parallaxes are shown in blue.  The M0--M6 field star sequence from
  \citet{Pecaut:2013} is shown in a red dashed line while the M6--L8 field
  sequence from \citet{Faherty:2016} is shown in a red dotted line. Young objects (aqua circles) lie
  above the sequence; their names are not on the plot to avoid crowding, but
  they are described in Section \ref{sec:young}. Known
  binaries (gray/yellow cirlces) generally lie above the field sequence and are
  listed in Section \ref{sec:outliers}. Five other sources that are too
  red for their spectral types and/or overluminous are shown with pink
  circles and are also discussed in Section \ref{sec:outliers}. \label{fig:colormag}}
\end{figure}

There are also 38 targets in Table \ref{tab:results} with no previous
trigonometric parallax including 7 stars with spectral types later than M8.  A
color-magnitude diagram for all the stars in our sample is shown in Figure
\ref{fig:colormag}. As expected, most of the new nearby objects have the expected
brightnesses and colors of old, field objects. Exceptions are discussed below.

\subsection{Discrepant Sources}

\begin{figure}[htb]
\includegraphics[height=2.4in,clip=1]{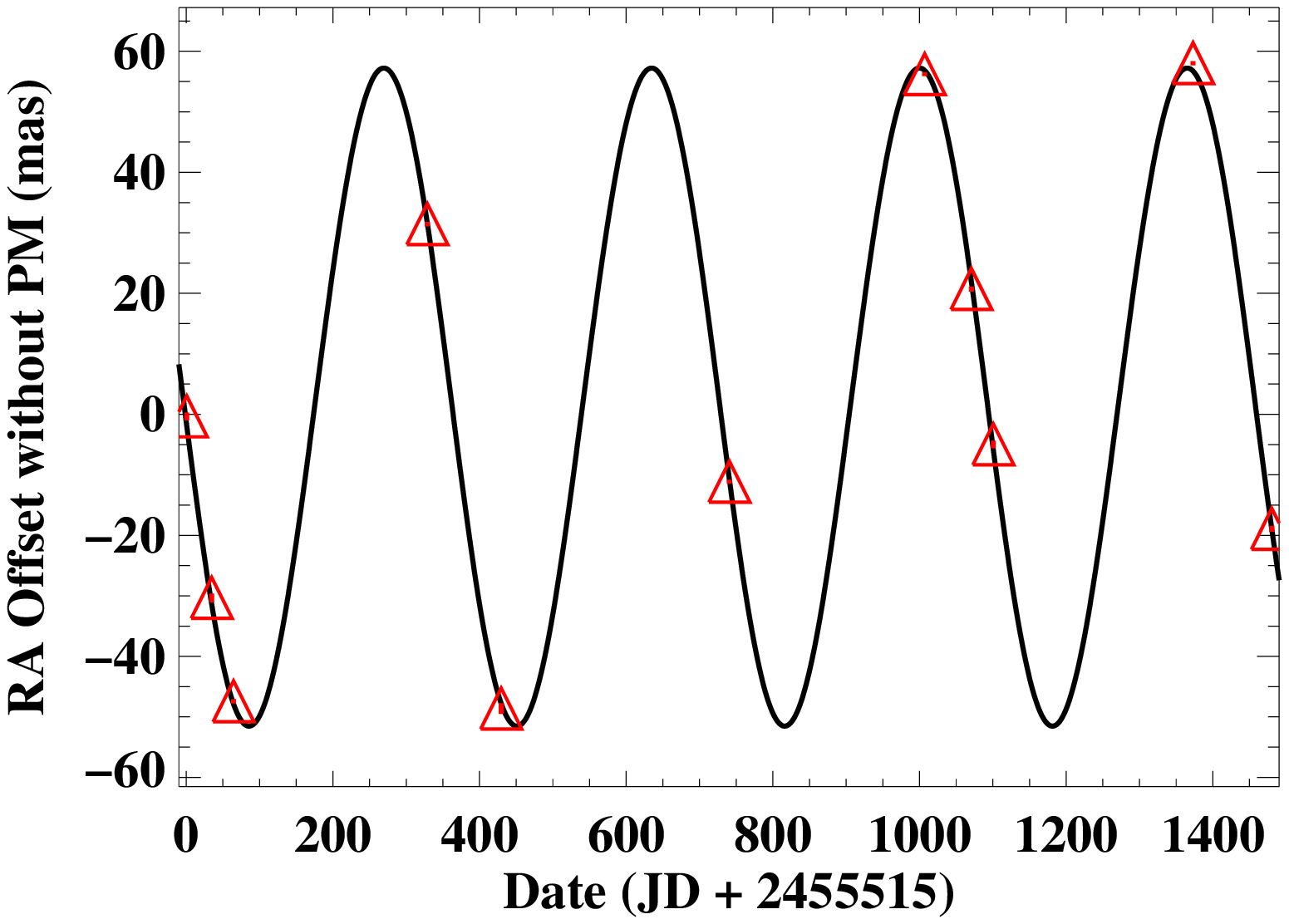}
\includegraphics[height=2.4in,clip=1]{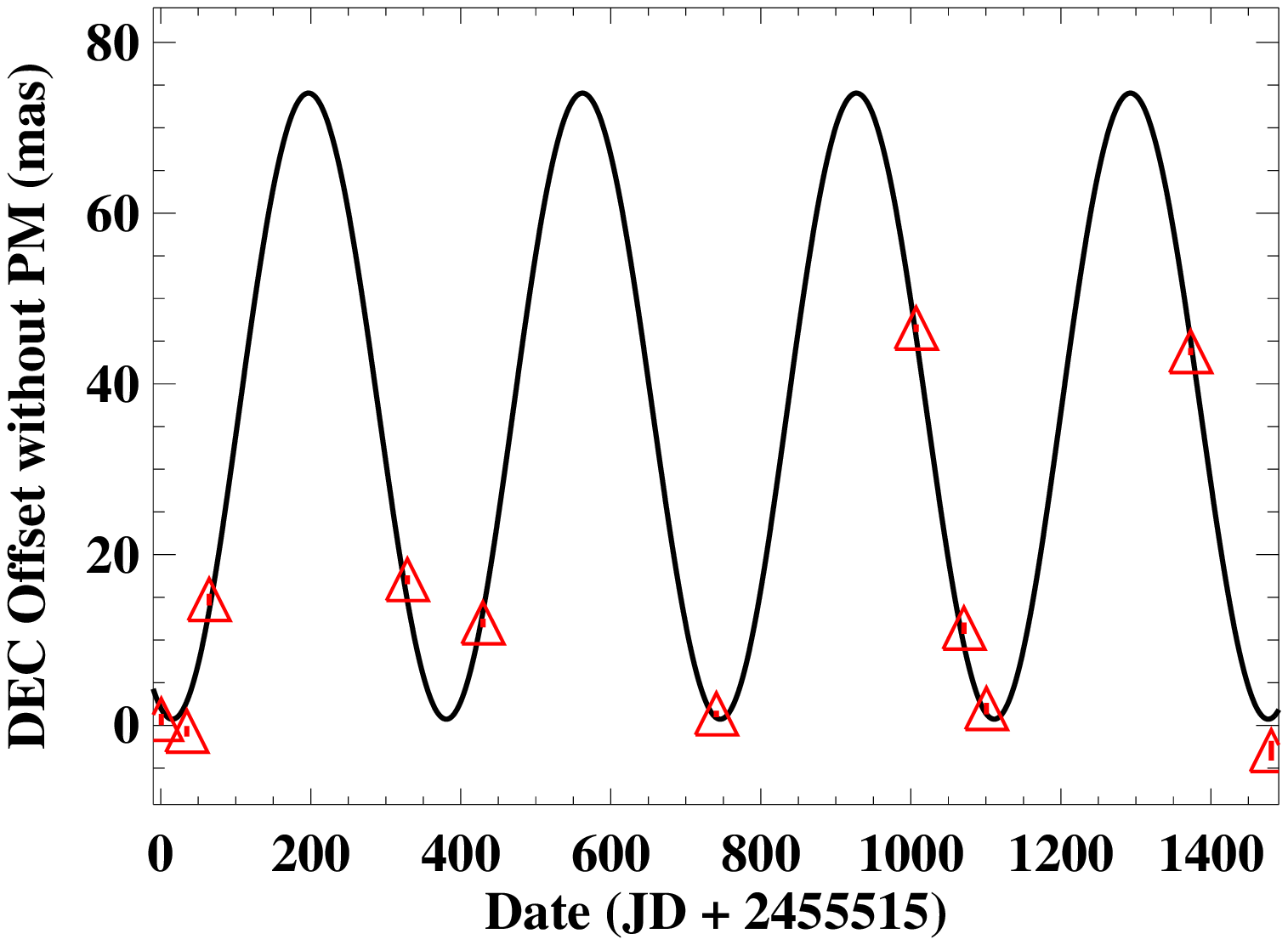}
\caption{Measurements of the motion of GJ 3198 in right ascension and
  declination after removing the proper motion, which otherwise dominates the
  scale of the plot. In all the plots shown, the position of the star in the
  first epoch of observation is taken to be (0,0). Typical per epoch
  uncertainties are $<$1 mas and are plotted but not usually visible within
  the symbols. The best fit parallax is shown with the solid
  line. \label{fig:GJ3198}}
\end{figure}

{\bf GJ 3198:} The literature value from \citet{Riedel:2010} is 67.3 $\pm$ 1.2
and our value is 57.2 $\pm$ 1.4. However, our (relative) proper motions agree
well: theirs is (483, -486) mas~yr$^{-1}$ and ours is (480, -474)
mas~yr$^{-1}$ . They have a baseline of 5.3 yr and we have a baseline of 4.1
yr.  Our parallax fits are shown in Figure \ref{fig:GJ3198}. So, the source of the
parallax discrepancy is unclear, but our parallax factor coverage is very
good, particularly in right ascension. With either parallax, the star's
position on the color-magnitude diagram is slightly too red for its absolute magnitude
and suggests probably binarity, but the star does not quite make the cuts we
impose to find such objects in Section \ref{sec:outliers}.

\begin{figure}
\includegraphics[height=2.4in,clip=1]{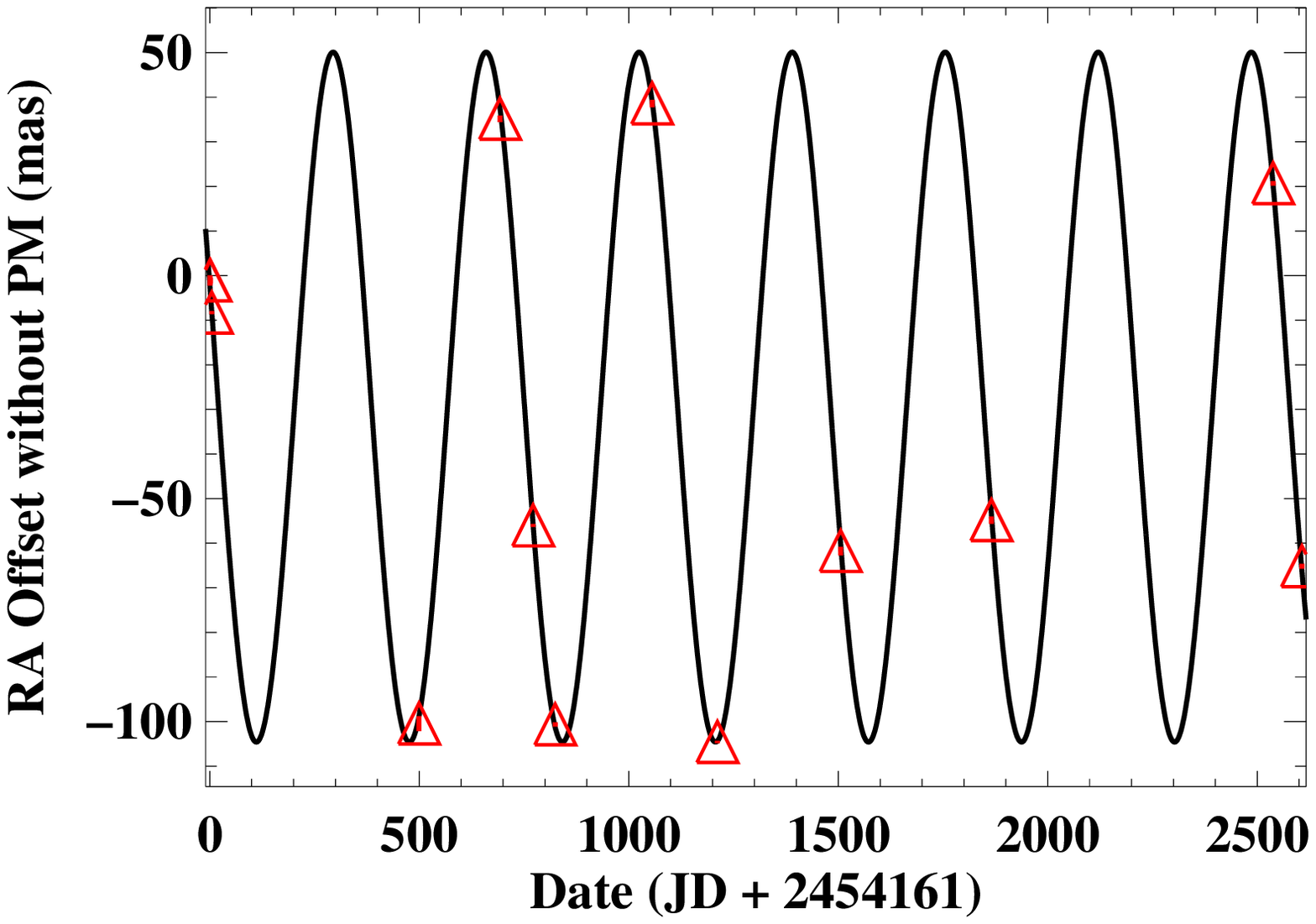}
\includegraphics[height=2.4in,clip=1]{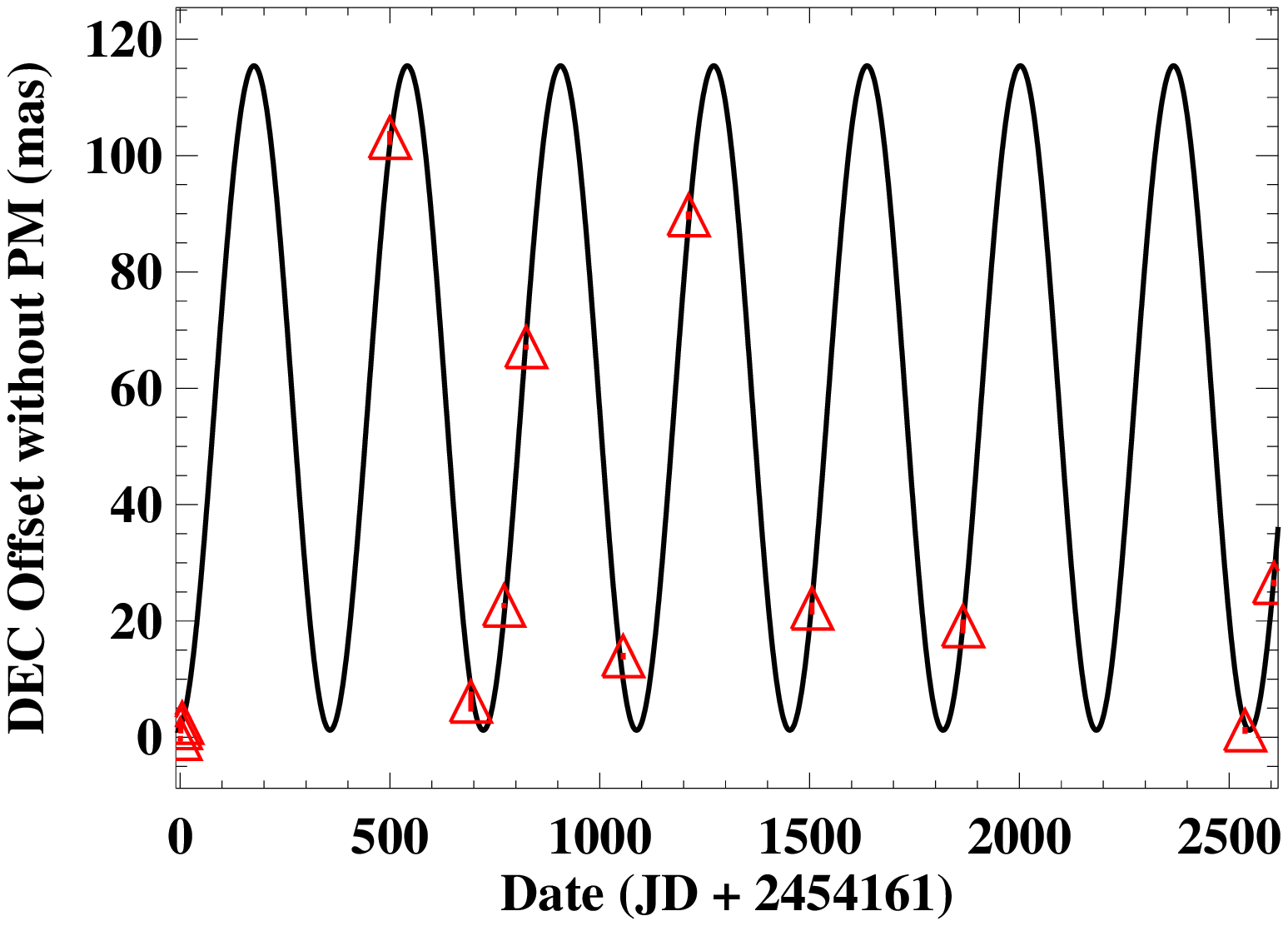}
\caption{Measurements of the motion of 2MASS J11553952-3727350 in right
  ascension and declination after removing the proper motion, which otherwise
  dominates the scale of the plot. The best fit
  parallax is shown with the solid line. \label{fig:2M1155}}
\end{figure}

{\bf 2MASS J11553952$−$3727350:} Our parallax of 84.4 $\pm$ 0.8 is 20\% smaller
than that of \citet{Faherty:2012}'s 104.4 $\pm$ 4.7.  Again, our relative
proper motions agree well: ours is (53.7, -784.49) and theirs is (66.8,
-777.9). They had a baseline of  2.5 yr and we have baseline of 7.1 yr.
Our parallax fits are shown in Figure \ref{fig:2M1155}. 

\begin{figure}[htb]
\includegraphics[height=2.4in,clip=1]{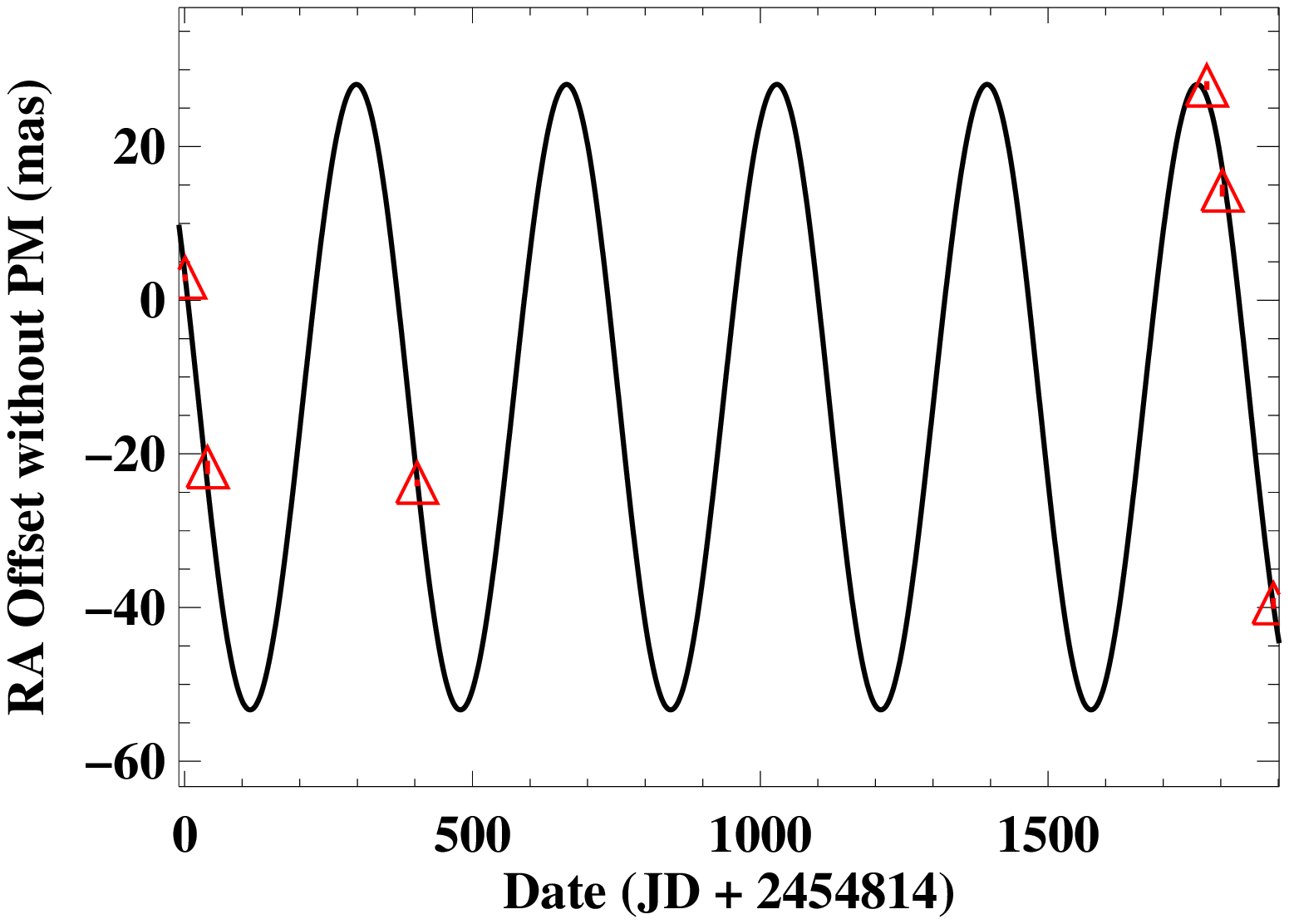}
\includegraphics[height=2.4in,clip=1]{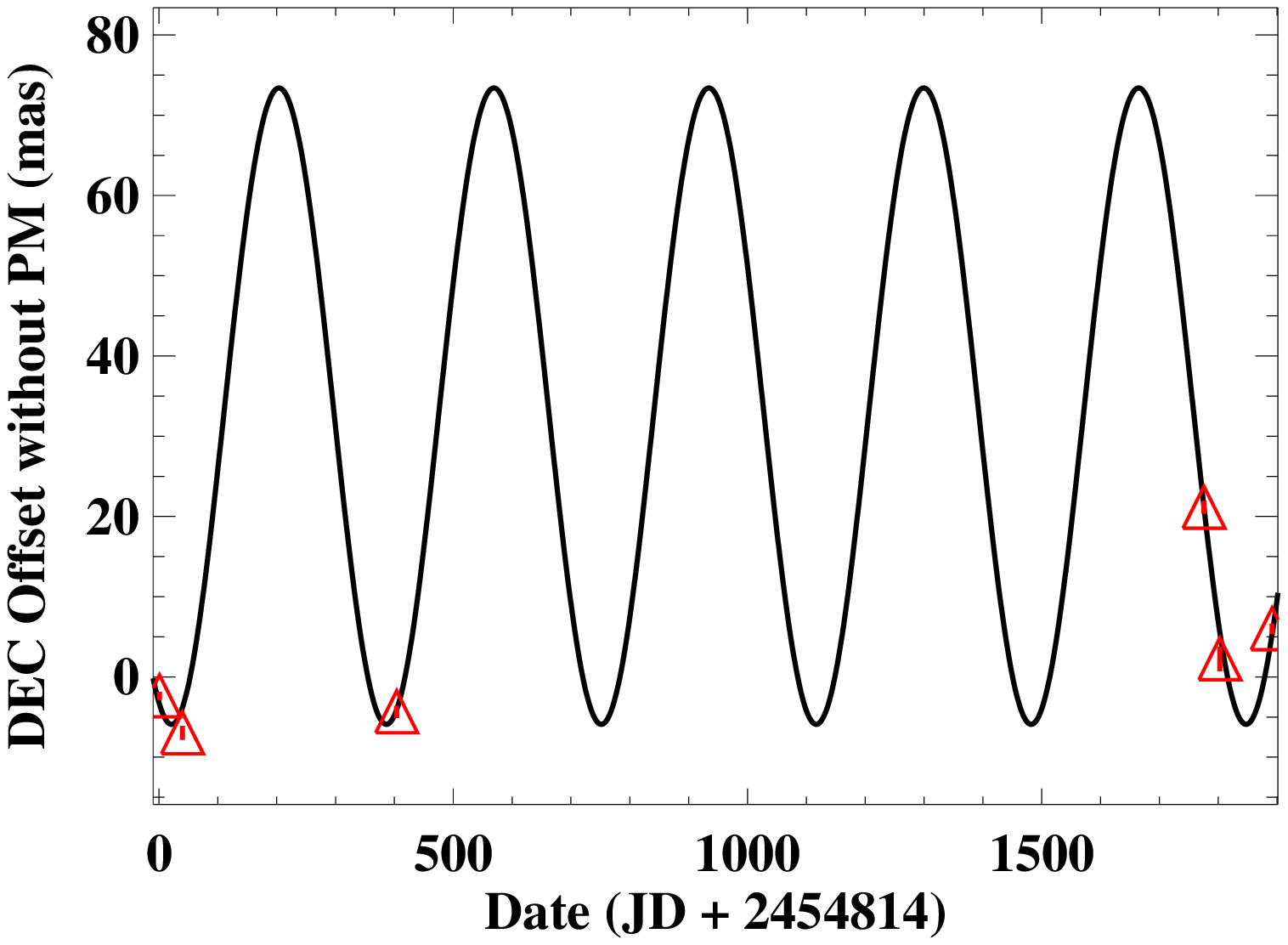}
\caption{Measurements of the motion of ESO 207-61 in
  ascension and declination after removing the proper motion, which otherwise
  dominates the scale of the plot. The best fit
  parallax is shown with the solid line. \label{fig:Ruiz207}}
\end{figure}

{\bf Ruiz (ESO) 207-61:} In the table, we gave the average of three literature
parallaxes, i.e. 54.7 mas \citep{Ianna:1995,Tinney:1996,vanaltena:1995}, but
the measured values range from 50.4 to 66.1 mas. We get 41.0 $\pm$ 1.6 mas. We
have dropped this source from our program, so we only have 6 epochs, but they
are spread over 5.2 yr with good coverage of the parallax factor. Our parallax
fits are shown in Figure \ref{fig:Ruiz207}.

\begin{figure}[htb]
\includegraphics[height=2.4in,clip=1]{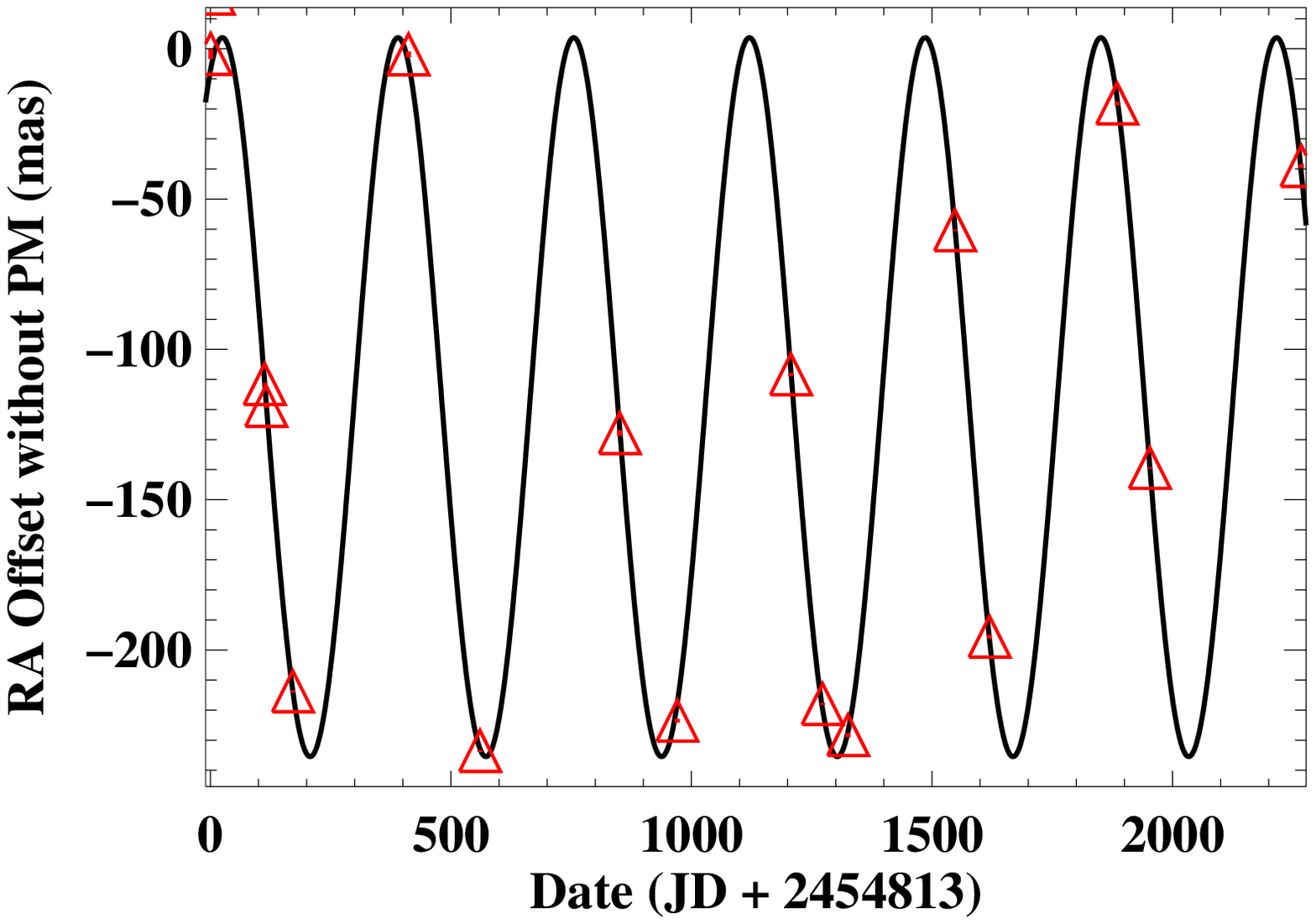}
\includegraphics[height=2.4in,clip=1]{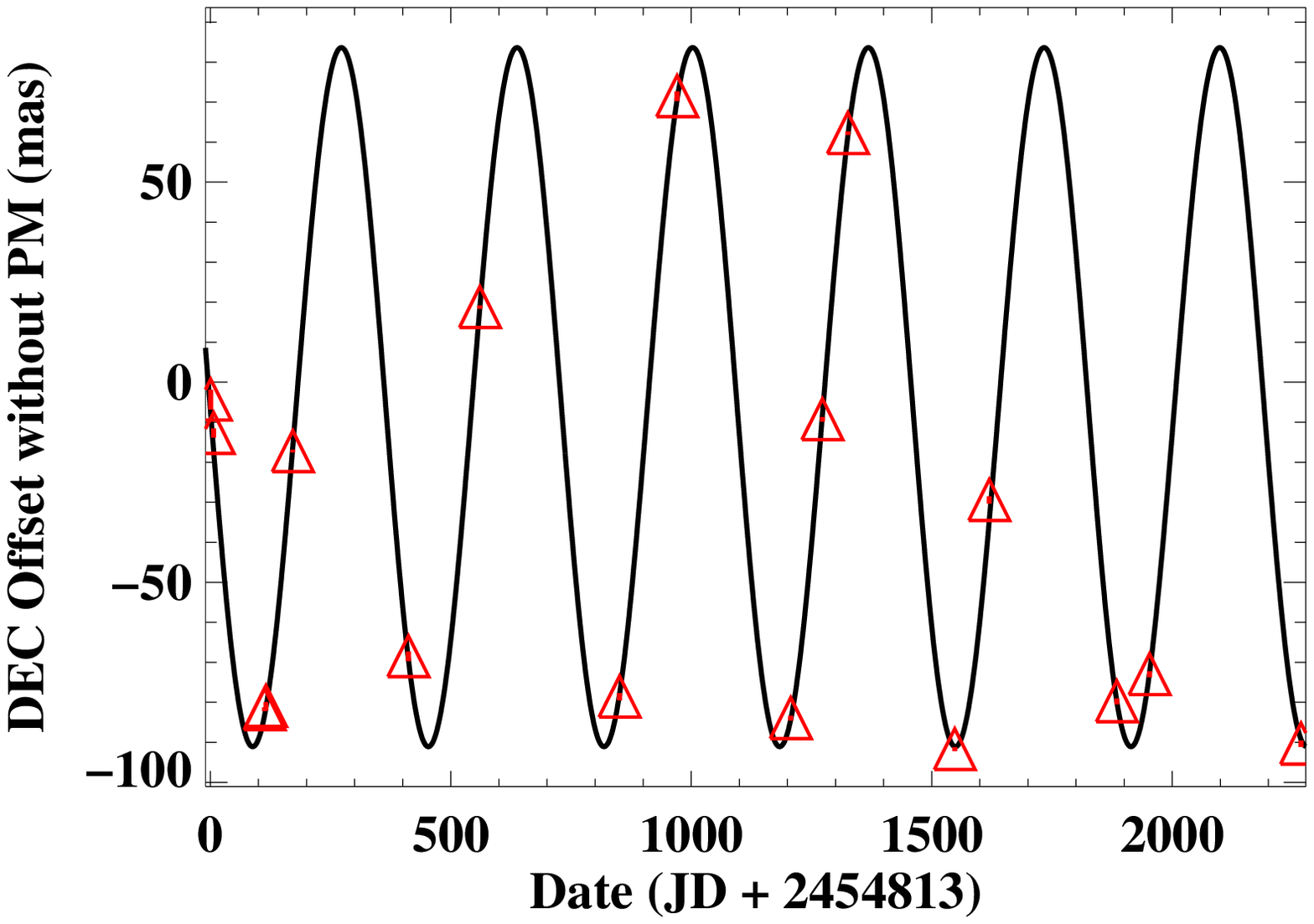}
\caption{Measurements of the motion of 2MASS J12590470-4336243 in right
  ascension and declination after removing the proper motion, which otherwise
  dominates the scale of the plot. The best fit
  parallax is shown with the solid line. \label{fig:J1259}}
\end{figure}

{\bf 2MASS J12590470$-$4336243: \label{s:J1259}} \citet{Deacon:2005} found a
parallax of 276 $\pm$ 41 mas for this object, which they refer to as
SIPS1259-4336, based on scanned UKST and ESO plates. They noted that their
derived distance (3.6 pc) made the object have an absolute magnitude too
bright for a single dwarf, and suggested it could be a binary. However, we
find a parallax of 129 mas, which puts the object twice as far away, at 7.8
pc, so it need not be a binary, and its absolute magnitude M$_J$=11.09
$\pm$0.05 is consistent with its color of J-W1=1.30 $\pm$ 0.03 for a single
M8. Our proper motion (not adjusted from the apparent value) of 1101.5 $\pm$
1.1 mas~yr$^{-1}$ in RA and $-$253.28 $\pm$ 0.30 mas~yr$^{-1}$ in DEC agrees
quite well with that of Deacon et al.: 1105 $\pm$ 4 mas~yr$^{-1}$ and $-$262
$\pm$ 4 mas~yr$^{-1}$ in RA and DEC respectively.   The parallax solution is shown
in Figure \ref{fig:J1259}.

\subsection{Notes on Interesting Individual Sources \label{sec:individual}}

{\bf 2MASS J01365662+0933473:} This nearby brown dwarf, type T2.5, is a
benchmark for the study of atmospheric variability and clouds in cool objects
\citep{Artigau:2009}. It had no previously published
parallax. \citet{Artigau:2006} found a photometric distance of 6.4 $\pm$ 0.3
pc, and our parallax gives a distance consistent with this, namely 6.14 $\pm$ 0.04
pc.

{\bf 2MASS J01392170$-$3936088}: The photometric distance to this source
computed in \citet{Deacon:2007} is 14.99$^{+8.96}_{-5.61}$ pc. Our
trigonometric parallactic distance is 8.80$\pm$0.04 pc, and the location of
the star in the M$_J - \rm (J-W1)$ color-magnitude diagram (Figure
\ref{fig:colormag}) does not look unusual. This is now added to the list of
stars within 10 pc.

{\bf LP 944-20:} This is a low-gravity, i.e. likely young, brown dwarf that is
not co-moving with a known young association \citep{Faherty:2016}.  Our
parallax of 154.4 $\pm$ 0.60 mas confirms the parallax measurement 155.9 $\pm$
1.0 mas of \citet{Dieterich:2014} that is markedly different from that of
\citet{Tinney:1996} (201.4 $\pm$ 4.2 mas).

{\bf GJ 3470:} This nearby M dwarf has a Neptune mass planet detected by
radial velocity and transit observations \citep{Bonfils:2012}. It has no
previously published trigonometric parallax; we get 34.15 $\pm$
0.66 mas or 29.28$^{+0.58}_{-0.56}$ pc.

The inferred planetary mass and radius depend sensitively on the stellar
properties. \citet{Demory:2013} measured a stellar density $\rho_\star =
2.91^{+0.37}_{-0.33} \rho_\odot$ and inferred M$_\star$=0.539$^{+0.047}_{-0.043}$ M$_\odot$,
R$_\star$=0.568$^{+0.037}_{-0.031}$ R$_\odot$, and distance=30.7$^{+2.1}_{-1.7}$ pc.

Our new distance is within their uncertainties, but we recompute the best
stellar mass and radius with a Monte Carlo that uses our distance and the
published photometry. Because of the density measurement, there are two nearly
independent methods to find R$_\star$. First, the physical size can be
determined from combining our distance with the K-band magnitude, via the
angular size relation of \citet{Kervella:2004}. Second, the stellar mass can be
determined from the V, J, H and K-band relations of \citet{Delfosse:2000} and
combined with the measured $\rho_\star$ of \citet{Demory:2013} to determine
R$_\star$.  We use a Monte Carlo to find the probability densities for both
independent estimates and then multiply the probability densities to get the
combined best estimate and its uncertainty: R$_\star$=0.550 $\pm$ 0.012 R$_\odot$.

Our best stellar radius is again within the uncertainties of the estimate of
\citet{Demory:2013}, but 3.2\% smaller on the mean and with smaller
uncertainty. This also reduces the inferred radius of the planet by the same
amount, and increases the planetary density by 10\% to 0.79 g~cm$^{-3}$.

{\bf 2MASS J20282035+0052265:} This is an L-dwarf binary system that was not
resolved in HST/NICMOS observations analyzed in \citet{Reid:2008} but was
resolved using new analysis techniques of the same data in \citet{Pope:2013}. The latter
work found it to be a nearly equal spectral type binary (L3+L4) and estimated
a new spectrophotometric distance of 26.1 $\pm$ 3.9 pc. Our parallax places
the binary at 30.1 $\pm$ 1.2 pc. We only have four epochs of data so we cannot
say if we observe orbital motion in the astrometric signal; it was dropped
from the planet search program for being too far away.

\subsubsection{Young Sources \label{sec:young}}

Stars can appear overluminous in the color-magnitude diagram
(e.g. Fig. \ref{fig:colormag}) because of youth. The companion to Fomalhaut,
LP 876-10 \citep{Mamajek:2013}, and AP Col \citep{Riedel:2011} are two examples
in Figure \ref{fig:colormag}. We also find two others.

{\bf G 161-71}: The spectrophotometric distance to this source is typically
given as $\sim$6.7 pc \citep{Reid:2002a,Scholz:2005a,Riaz:2006}, but our
parallactic distance is 13.26 $\pm$ 0.14 pc. \citet{Malo:2014} measured an RV
of 13.5 $\pm$ 0.4 km~s$^{-1}$ and listed it as a possible Argus association
member.  Using our parallax and proper motions combined with this RV, we
confirm a 99.99\% probability of membership in the 30-50 Myr old Argus
association using the BANYAN I tool \citep{Malo:2013}. An overluminosity of
1.5 mag is possible for such a young star \citep{Gagne:2015}. In addition, the
enhanced X-ray luminosity of this star \citep{Riaz:2006} is also consistent
with that of other young stars \citep{Shkolnik:2009}.

{\bf LP 870-65}: This is an M4 or M4.5 star with a spectrophotometric distance
in \citet{Scholz:2005a} of 8.7 pc, and our parallactic distance is 18.22 $\pm$
0.19 pc.  Indeed, \citet{Bowler:2015} identified this star, also known as NLTT
48651, as young based on its X-ray and UV luminosity. That paper also gives a
radial velocity of -7.5 $\pm$ 0.7 (E. Shkolnik, personal communication) and
suggests a tentative association with the AB Dor moving group. Using our
parallax and proper motions combined with this RV, the BANYAN I tool confirms
a 100\% probability of membership in the $\sim$100 Myr old AB Dor Association.

\subsubsection{Overluminous and/or Red Sources \label{sec:outliers}}

{\bf Binaries:} Several known binaries are in our sample; those that are equal
brightness appear overluminous in Figure \ref{fig:colormag}: GJ 2005
\citep{Leinert:1994}, Kelu-1 \citep{Gelino:2006}, G 124-62B \citet{Bouy:2003},
GJ 3900 \citep{Bonfils:2013}, GJ 4074 \citep{Bonfils:2013}, LP 869-19
  \citep{Malo:2014}, 2MASS
J20282035+0052265 \citep{Pope:2013}, and $\epsilon$ Indi B
\citep{McCaughrean:2004}. The companions to 2MASS J04234858-0414035
\citep{Burgasser:2005} and 2MASS J13153094-2649513 \citep{Burgasser:2011} are T
dwarfs and do not cause noticable overluminosity.  Suprisingly, 2MASS
J02052940-1159296 \citep{Koerner:1999}, which is an equal flux ratio binary,
does not look overluminous.

In addition to these known binaries, we search for stars that appear
overluminous or redder than expected based on their spectral types. For M0 - M6
spectral types, we search for stars that lie redder than the field sequence as
given in \citet{Pecaut:2013}\footnote[1]{Updated at
  \url{http://www.pas.rochester.edu/$\sim$emamajek/EEM\_dwarf\_UBVIJHK\_colors\_Teff.txt}}
by more than the combined 1$\sigma$ uncertainties in the dwarf sequence and the
stars' individual color uncertainties. Since Pecaut \& Mamajek do not provide
uncertainties on the colors, we computed J$-$W1  for $\sim$20 stars in each spectral type
bin taken from DwarfArchives.org\footnote[2]{List of M dwarfs at
  http://spider.ipac.caltech.edu/staff/davy/ARCHIVE/index.shtml}. For M4, M5 and
M6 stars, we find a color and dispersion of 1.00 $\pm$ 0.06, 1.12 $\pm$ 0.08,
and 1.16 $\pm$ 0.08 mag respetively.  We assume a 0.08 mag uncertainty for M0-M3
also. For M7 and later spectral types, we search for stars that lie above the
field sequence given in \citet{Faherty:2016}.  Combined, we find four stars that
appear overluminous or redder than expected: DY Psc, GJ 1123, GJ 1129, and 2MASS
J16184503-1321297.

These stars are peculiar. In principle, they could be candidate young
stars. All of these stars have absolute magnitudes that are more than 0.75 mag
from their expected values based on their J$-$W1, so are not just obviously
equal brightness binaries.  DY Psc is particularly red (J$-$W1 = 1.82 $\pm$
0.04) for its optically determined spectral type of M9.5 (J$-$W1 =
1.5). However, none of these stars has X-ray emission detected in the ROSAT
all-sky survey \citep{Boller:2016} or were strong UV emitters, at the level of
the known young stars, in the GALEX survey.

\section{Discussion and Summary}

Parallaxes combined with infrared colors can identify interesting low mass stars
and brown dwarfs that are young and/or in multiple systems. Only four of the
targets in our sample are in the Tycho-2 catalog and would therefore be expected
to have full astrometric solutions including parallax in the first GAIA data
release in 2016. These are GJ 3379 (M4), G 108-21 (M3.5), GL 452.1 (M4.5), and
LTT 7434 (M4).

Thirty-two of the stars here are not part of our long-term
monitoring program for any of a number of reasons including: being too far away
($\pi < $50 mas), being a close visual binary or stellar spectroscopic binary,
or having a bad astrometric reference frame. We are continuing to observe all
the stars that have more than 10 epochs in Table \ref{tab:results}.

\acknowledgements 

We thank the staff of Las Campanas Observatory for their ongoing support of
this long-term program.  Jackie Faherty and Jonathan Gagne provided helpful
input. This work has been supported in part by NSF grant AST-0352912, NASA
Origins of Solar Systems grant NNX09AF62G, and NASA Astrobiology Institute
grant NNA09DA81A. This research has made use of the SIMBAD and Vizier
databases, operated at CDS, Strasbourg, France. This publication makes use of
data products from the Two Micron All Sky Survey, which is a joint project of
the University of Massachusetts and the Infrared Processing and Analysis
Center/California Institute of Technology, funded by the National Aeronautics
and Space Administration and the National Science Foundation.  This
publication makes use of data products from the Wide-field Infrared Survey
Explorer, which is a joint project of the University of California, Los
Angeles, and the Jet Propulsion Laboratory/California Institute of Technology,
funded by the National Aeronautics and Space Administration. This publication
made use of the Mikulski Archive for Space Telescopes (MAST). STScI is
operated by the Association of Universities for Research in Astronomy, Inc.,
under NASA contract NAS5-26555. Support for MAST for non-HST data is provided
by the NASA Office of Space Science via grant NNX09AF08G and by other grants
and contracts.

\bibliographystyle{apj}
\bibliography{capspar_format}

\defcitealias{Reid:1995}{1}
\defcitealias{Faherty:2009}{2}
\defcitealias{Kendall:2007}{3}
\defcitealias{PhanBao:2006}{4}
\defcitealias{Hawley:1996}{5}
\defcitealias{Cruz:2002}{6}
\defcitealias{Reid:2003}{7}
\defcitealias{Riaz:2006}{8}
\defcitealias{Bonfils:2013}{9}
\defcitealias{Bowler:2010}{10}
\defcitealias{Crifo:2005}{11}
\defcitealias{Reid:2005}{12}
\defcitealias{Marshall:2008}{13}
\defcitealias{Scholz:2005a}{14}
\defcitealias{Lodieu:2005}{15}
\defcitealias{Reid:2007}{16}

\begin{deluxetable}{lllllcccclllrrrr}
\tablecaption{Observational information for CAPSCam targets with solved parallaxes\label{tab:obs}}
\tabletypesize{\scriptsize}
\rotate
\tablewidth{0pt}
\tablehead{
\colhead{Name} & \colhead{RA}  &\colhead{Decl.}
                                          &\colhead{Sp.}
                                            &\colhead{Ref}
&\colhead{t$_{\rm int,FF}$} &\colhead{t$_{\rm int,GW}$}
                                                  &\colhead{\#} 
&\colhead{\#}
&\colhead{Date$_{start}$} &\colhead{Date$_{end}$} &\colhead{$\Delta$ t}
&\colhead{m$_{J}$} &\colhead{$\sigma$m$_{J}$}
&\colhead{m$_{W1}$} &\colhead{$\sigma$m$_{W1}$}\\

\colhead{} & \colhead{hh mm ss.ss}  &\colhead{dd mm ss.s}
                                          &\colhead{type}
                                             &\colhead{}
&\colhead{(s)} &\colhead{(s)}
                                                  &\colhead{ref $\star$s}
&\colhead{epochs}
&\colhead{(JD)} &\colhead{(JD)} &\colhead{(yr)}
&\colhead{mag} &\colhead{mag}
&\colhead{mag} &\colhead{mag}
}
\startdata
GJ 1002       &00 06 43.25 & -07 32 14.7    &M5.5    &\citetalias{Reid:1995}    & 120   &  2           & 13  & 14   & 2455141.6 & 2456999.5 & 5.1  & 8.323   & 0.019   & 7.16    & 0.05 \\
LEHPM 193     &00 07 07.80 & -24 58 03.8    &M7      &\citetalias{Faherty:2009} & 120   &  \nodata     & 13  & 5    & 2455076.7 & 2456492.8 & 3.9  & 13.115  & 0.024   & 11.84   & 0.02 \\
DY Psc        &00 24 24.63 & -01 58 20.1    &M9.5    &\citetalias{Faherty:2009} & 120   &  30          & 35  & 16   & 2454722.7 & 2456995.5 & 6.2  & 11.992  & 0.035   & 10.17   & 0.02 \\
GJ 2005 A     &00 24 44.18 & -27 08 25.2    &M5.5    &\citetalias{Reid:1995}    & 120   &  5           & 22  & 5    & 2455410.9 & 2456587.7 & 3.2  & 9.254   & 0.034   & 7.81    & 0.02 \\
LEHPM 1130    &00 58 06.43 & -53 18 09.2    &        &                          & 120   &  \nodata     & 20  & 5    & 2454347.8 & 2456491.8 & 5.9  & 12.998  & 0.024   & 11.73   & 0.02 \\
GJ 1028       &01 04 53.68 & -18 07 29.3    &M5      &\citetalias{Reid:1995}    & 120   &  1           & 23  & 20   & 2454346.8 & 2456993.5 & 7.2  & 9.387   & 0.026   & 8.22    & 0.02 \\
LP 647-13     &01 09 51.17 & -03 43 26.4    &M9      &\citetalias{Faherty:2009} & 120   &  20          & 17  & 10   & 2454290.9 & 2456488.9 & 6.0  & 11.694  & 0.021   & 10.08   & 0.02 \\
LEHPM 1563    &01 27 31.96 & -31 40 03.2    &M7      &\citetalias{Kendall:2007} & 120   &  30          & 19  & 7    & 2454343.8 & 2456554.8 & 6.1  & 12.659  & 0.026   & 11.4    & 0.02 \\
\enddata
\tablecomments{Spectral types are visual wavelength where available. J
  magnitudes are from 2MASS, W1 magnitudes are from ALLWISE. \\
(Table \ref{tab:obs} will be published online in its entirety in a machine readable
  format. The published paper will show a stub table for guidance regarding its form and
  content.)}
\tablerefs{
[\citetalias{Reid:1995}] \citet{Reid:1995};
[\citetalias{Faherty:2009}] \citet{Faherty:2009};
[\citetalias{Kendall:2007}] \citet{Kendall:2007};
[\citetalias{PhanBao:2006}] \citet{PhanBao:2006};
[\citetalias{Hawley:1996}] \citet{Hawley:1996};
[\citetalias{Cruz:2002}] \citet{Cruz:2002};
[\citetalias{Reid:2003}] \citet{Reid:2003};
[\citetalias{Riaz:2006}] \citet{Riaz:2006};
[\citetalias{Bonfils:2013}] \citet{Bonfils:2013};
[\citetalias{Bowler:2010}] \citet{Bowler:2010};
[\citetalias{Crifo:2005}] \citet{Crifo:2005};
[\citetalias{Reid:2005}] \citet{Reid:2005};
[\citetalias{Marshall:2008}] \citet{Marshall:2008};
[\citetalias{Scholz:2005a}] \citet{Scholz:2005a};
[\citetalias{Lodieu:2005}] \citet{Lodieu:2005};
[\citetalias{Reid:2007}] \citet{Reid:2007}
}
\end{deluxetable}

\defcitealias{vanaltena:1995}{1}
\defcitealias{Tinney:1995}{2}
\defcitealias{Costa:2005}{3}
\defcitealias{Harrington:1980}{4}	
\defcitealias{Dahn:1988}{5}
\defcitealias{Dahn:2002}{6}
\defcitealias{Dahn:1982}{7}
\defcitealias{Henry:2006}{8}
\defcitealias{Costa:2006}{9}
\defcitealias{Riedel:2010}{10}
\defcitealias{Faherty:2012}{11}
\defcitealias{Tinney:1996}{12}
\defcitealias{Dieterich:2014}{13}
\defcitealias{Vrba:2004}{14}
\defcitealias{Riedel:2014}{15}
\defcitealias{Marocco:2013}{16}
\defcitealias{Gatewood:2008}{17}
\defcitealias{Riedel:2011}{18}
\defcitealias{Andrei:2011}{19}
\defcitealias{Heintz:1994}{20}
\defcitealias{Ianna:1995}{21}
\defcitealias{Anglada:2012}{22}
\defcitealias{Smart:2010}{23}
\defcitealias{Jao:2005}{24}
\defcitealias{Jao:2011}{25}
\defcitealias{Deacon:2001}{26}
\defcitealias{vanleeuwen:2007}{27}
\defcitealias{Harrington:1993}{28}
\defcitealias{Deacon:2005}{29}
\defcitealias{Dupuy:2012}{30}
\defcitealias{Pravdo:2009}{31}
\defcitealias{Mamajek:2013}{32}

\begin{deluxetable}{lllrrrrrrrrlrrrr}
\tablecaption{Astrometric Results\label{tab:results}}
\tabletypesize{\scriptsize}
\rotate
\tablewidth{0pt}
\tablehead{
\colhead{Name}                &\colhead{RA}                   &\colhead{Decl.}
&\colhead{$\pi_{\rm rel}$}      &\colhead{$\sigma \pi_{\rm rel}$}
&\colhead{Zero}               &\colhead{$\sigma_{\rm ZPt}$}
&\colhead{$\pi_{\rm Lit.}$}  &\colhead{$\sigma \pi_{\rm Lit}$} &\colhead{Ref}
&\colhead{$\mu_{RA}$}          &\colhead{$\sigma\mu$}
&\colhead{$\mu_\delta$}        &\colhead{$\sigma\mu$}
&\colhead{$\pi_{\rm abs}$}      &\colhead{$\sigma \pi_{\rm abs}$}\\
\colhead{}                    &\colhead{} &\colhead{}
&\colhead{(mas)}              &\colhead{(mas)}
&\colhead{Pt}                 &\colhead{(mas)}
&\colhead{(mas)}              &\colhead{(mas)}                &\colhead{}
&\colhead{($cos\delta$,rel)}  &\colhead{($cos\delta$,rel)}
&\colhead{(rel)}              &\colhead{(rel)}
&\colhead{(mas)}              &\colhead{(mas)}\\
\colhead{}                    &\colhead{}                     &\colhead{}
&\colhead{}                   &\colhead{}
&\colhead{(mas)}              &\colhead{}
&\colhead{}                   &\colhead{}                     &\colhead{}
&\colhead{(mas}               &\colhead{(mas}
&\colhead{(mas}               &\colhead{(mas}
& \colhead{}                  &\colhead{}
\\
\colhead{}                    &\colhead{}                    &\colhead{} 
&\colhead{}                   &\colhead{} 
&\colhead{}                   &\colhead{}
&\colhead{}                   &\colhead{}                    &\colhead{}
&\colhead{yr$^{-1}$)}          &\colhead{yr$^{-1}$)} 
&\colhead{yr$^{-1}$)}          &\colhead{yr$^{-1}$)}
&\colhead{}                   &\colhead{}
}
\startdata
GJ 1002       &00 06 43.25 & -07 32 14.7  & 206.92 & 0.97   & -0.50  & 0.75   & 213    &  3.6     &\citetalias{vanaltena:1995}                                    & -805.16   & 0.56 & -1870.61      & 0.37                    & 207.42 & 1.23\\
LEHPM 193     &00 07 07.80 & -24 58 03.8  & 38.80  & 1.27   & 0.00   & 0.40   &        &          &                                                               & 184.84    & 0.66 & -56.55        & 0.52                    & 38.80  & 1.33\\
DY Psc        &00 24 24.63 & -01 58 20.1  & 79.78  & 0.91   & -0.93  & 0.26   & 84.3   &  2.6     &\citetalias{vanaltena:1995},\citetalias{Tinney:1995}           & -78.23    & 0.30 & 141.47        & 0.60                    & 80.71  & 0.95\\
GJ 2005 A     &00 24 44.18 & -27 08 25.2  & 123.97 & 11.40  & -8.33  & 0.03   &129.7   & 2.4      &\citetalias{vanaltena:1995},\citetalias{Costa:2005}            & -106.57   & 5.72 & 690.64        & 7.83                    & 132.3  & 11.4\\
LEHPM 1130    &00 58 06.43 & -53 18 09.2  & 28.39  & 0.97   & 0.00   & 0.40   &        &          &                                                               & -202.64   & 0.43 & -237.61       & 0.22                    & 28.39  & 1.05\\
GJ 1028       &01 04 53.68 & -18 07 29.3  & 101.43 & 0.43   & -0.80  & 0.75   & 99.80  &  5.00    &\citetalias{vanaltena:1995}                                    & 1274.41   & 0.12 & 494.08        & 0.35                    & 102.23 & 0.86\\
LP 647-13     &01 09 51.17 & -03 43 26.4  & 98.32  & 1.30   & 0.77   & 1.58   & 104.23 &  2.29    &\citetalias{Costa:2005}                                        & 365.34    & 0.42 & 15.57         & 0.42                    & 97.55  & 2.04\\
LEHPM 1563    &01 27 31.96 & -31 40 03.2  & 38.22  & 3.89   & -1.49  & 0.87   &        &          &                                                               & 284.10    & 1.48 & 137.70        & 8.91                    & 39.71  & 3.99\\
\enddata
\tablecomments{\ Literature parallaxes are weighted averages when more than one value exists\\
(Table \ref{tab:results} will be published online in its entirety in a machine
  readable format. The published paper will show a stub table for guidance
  regarding its form and content.)}
\tablerefs{
[\citetalias{vanaltena:1995}] \citet{vanaltena:1995};
[\citetalias{Tinney:1995}] \citet{Tinney:1995};
[\citetalias{Costa:2005}] \citet{Costa:2005};
[\citetalias{Harrington:1980}] 	\citet{Harrington:1980};
[\citetalias{Dahn:1988}] \citet{Dahn:1988};
[\citetalias{Dahn:2002}] \citet{Dahn:2002};
[\citetalias{Dahn:1982}] \citet{Dahn:1982};
[\citetalias{Henry:2006}] \citet{Henry:2006};
[\citetalias{Costa:2006}] \citet{Costa:2006};
[\citetalias{Riedel:2010}] \citet{Riedel:2010};
[\citetalias{Faherty:2012}] \citet{Faherty:2012};
[\citetalias{Tinney:1996}] \citet{Tinney:1996};
[\citetalias{Dieterich:2014}] \citet{Dieterich:2014};
[\citetalias{Vrba:2004}] \citet{Vrba:2004};
[\citetalias{Riedel:2014}] \citet{Riedel:2014};
[\citetalias{Marocco:2013}] \citet{Marocco:2013};
[\citetalias{Gatewood:2008}] \citet{Gatewood:2008};
[\citetalias{Riedel:2011}] \citet{Riedel:2011};
[\citetalias{Andrei:2011}] \citet{Andrei:2011};
[\citetalias{Heintz:1994}] \citet{Heintz:1994};
[\citetalias{Ianna:1995}] \citet{Ianna:1995};
[\citetalias{Anglada:2012}] \citet{Anglada:2012};
[\citetalias{Smart:2010}] \citet{Smart:2010};
[\citetalias{Jao:2005}] \citet{Jao:2005};
[\citetalias{Jao:2011}] \citet{Jao:2011};
[\citetalias{Deacon:2001}] \citet{Deacon:2001};
[\citetalias{vanleeuwen:2007}] \citet{vanleeuwen:2007};
[\citetalias{Harrington:1993}] \citet{Harrington:1993};
[\citetalias{Deacon:2005}] \citet{Deacon:2005};
[\citetalias{Dupuy:2012}] \citet{Dupuy:2012};
[\citetalias{Pravdo:2009}] \citet{Pravdo:2009};
[\citetalias{Mamajek:2013}] \citet{Mamajek:2013};
}
\end{deluxetable}                                                                                                                                                                                                 
\end{document}